\documentclass[global,twocolumn]{svjour} %,referee
\if@twocolumn
    \newenvironment{widetext}
        {%
            \begin{strip}
            \rule{\dimexpr(0.5\textwidth-0.5\columnsep-0.4pt)}{0.4pt}%
            \rule{0.4pt}{6pt}
            \par %\vspace{6pt}
            \parindent \@parindent
        }%
        {%
            \par
            \hfill\rule[-6pt]{0.4pt}{6.4pt}%
            \rule{\dimexpr(0.5\textwidth-0.5\columnsep-1pt)}{0.4pt}
            \end{strip}
        }
\else
    \newenvironment{widetext}{}{}
\fi

\usepackage{graphicx}
\usepackage{dsfont}
\usepackage{amsmath}
\usepackage{amssymb}
\newcommand{\1}{{\mathds{1}}}
\newcommand{\N}{{\mathds{N}}}
\newcommand{\Peven}{P_{\mathrm{even}}}
\newcommand{\Podd}{P_{\mathrm{odd}}}
\newcommand{\Psucc}{P_{\mathrm{succ}}}
\usepackage[latin1]{inputenc}
\usepackage{subfigure}
\newcommand{\Ket}[1]{\left| #1 \right\rangle}
\newcommand{\ket}[1]{| #1 \rangle}
\newcommand{\bra}[1]{\langle #1 |}

\journalname{Applied Physics B}

\begin{document}
\title{On the error analysis of quantum repeaters with encoding}
\subtitle{}%Do you have a subtitle?\\ If so, write it here
\author{Michael Epping\inst{1} \and Hermann Kampermann\inst{1} \and Dagmar Bru\ss{}\inst{1}% etc
}                     % Do not remove
\institute{Institut für Theoretische Physik III, Heinrich-Heine-Universität Düsseldorf, Germany}
\date{Received: date / Revised version: date}
\maketitle
\begin{abstract}
Losses of optical signals scale exponentially with the distance. Quantum repeaters are devices that tackle these losses in quantum communication by splitting the total distance into shorter parts. 
Today two types of quantum repeaters are subject of research in the field of quantum information: Those that use two-way communication and those that only use one-way communication. Here we explain the details of the performance analysis for repeaters of the second type. Furthermore we compare the two different schemes. Finally we show how the performance analysis generalizes to large-scale quantum networks.
\end{abstract}
\section{Introduction}
\label{sec:intro}
Signals in long distance telecommunications are subject to corruptions. Typically the amplitude decreases exponentially with the covered
distance~\cite{Gisin02}. Thus intermediate repeaters which amplify and purify the signal are necessary building blocks for reliable transmission. In
quantum cryptography and communication the signals transport coherent quantum information.\\
One possibility to overcome the exponential scaling of losses with distance is the entanglement swapping and \hbox{-}distillation based repeater
scheme, which was developed by H.-J. Briegel, W. Dür, J. Cirac and P. Zoller in \cite{Briegel98}. Here entangled pairs are distributed
amongst neighboring repeater stations and Bell measurements on each station result in entangled states covering a larger distance (so-called
entanglement swapping). These operations introduce errors which can be tackled by entanglement distillation, i.e. protocols that concentrate
several imperfect copies of entangled states into a single copy with higher fidelity with respect to a maximally entangled state~\cite{Bennett96,Deutsch96,Duer03}. Two-way
classical communication is used to acknowledge reception of photons and success of distillation.\\
A different approach, introduced by L. Jiang, J.  Taylor, K. Nemoto, W. Munro, R. Van Meter, and M. Lukin in \cite{Jiang09}, replaces the
entanglement distillation step by the use of quantum error correction codes for forward error correction, i.e. communication is
only required in one direction. In comparison to the previous schemes these improve the repeater rate at the cost of being more demanding in
terms of resources and the quality of operations. Subsequent work considered different codes and improved the error
analysis~\cite{Munro10,Fowler10,Munro12,Muralidharan14,Epping15}.\\
In the present paper we attempt to give a simple analysis of repeaters of the latter type. 
Before describing the error analysis we motivate the quantum repeater circuits in Section~\ref{sec:thecircuit}. We use the stabilizer formalism, which is very convenient in this context. Section~\ref{sec:errormodel} summarizes the error model of depolarizing noise, which is widely used in the context of error correction. Section~\ref{sec:repeatercircuit} then discusses a quantum repeater in the circuit model. We put emphasis on the sources of errors and their propagation and estimate the effective error rates of the physical qubits. The considered repeater schemes are based on error correction codes, which implies that several physical qubits form a logical qubit. This redundancy allows to correct errors and the strength of this correction is discussed in Section~\ref{sec:logicalerrorrates}. The overall performance of the repeater scheme, i.e. its ability to produce a specific entangled state, is then analyzed in Section~\ref{sec:finalstate} and compared to the original scheme. Finally we sketch in Section~\ref{sec:multipartite} how repeaters with encoding generalize to large-scale quantum networks using the ideas we presented in~\cite{Epping15}.
\section{The circuit of quantum repeaters can be understood in the stabilizer formalism}
\label{sec:thecircuit}
Before diving into the error analysis of the quantum repeater we motivate the circuit using the stabilizer formalism~\cite{Nielsen00,Gottesman97}. This language simplifies the multipartite generalization. Furthermore we think it is an aesthetic way of constructing and understanding the circuit.\\
A state $\ket{\psi}$ is said to be stabilized by an operator $s$ if it is an eigenstate of $s$ to the eigenvalue $+1$, i.e.
\begin{equation}
s\ket{\psi}=\ket{\psi}. 
\end{equation}
The set of all operators that stabilize the state is called the stabilizer of the state. An $n$ qubit state can be uniquely defined by $n$ independent operators $g_i$, $i=1,2,...,n$, that stabilize it. They generate the stabilizer of the state, i.e. the product of two stabilizer elements is contained in the stabilizer, too. Here we shortly call an element of the stabilizer, i.e. the operator, a stabilizer. No confusion should arise from this abbreviation.\\
\subsection{The stabilizer of a maximally entangled state}
Consider the maximally entangled state shared by parties $A$ and $B$
\begin{equation}
 \Ket{\includegraphics[width=7mm]{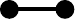}}_{AB}=\frac{1}{\sqrt{2}}(\ket{0}_A\ket{+}_B+\ket{1}_A\ket{-}_B),\label{eq:bipartitestate}
\end{equation}
where $\ket{0}$ and $\ket{1}$ form the canonical basis of the Hilbert space of a single qubit and $\ket{\pm}=\frac{1}{\sqrt{2}}(\ket{0}\pm \ket{1})$.
The state $\Ket{\includegraphics[width=7mm]{bellpair}}_{AB}$ is local unitary equivalent to any Bell pair in the standard notation,
\begin{align}
 \ket{\psi^+}=&\frac{1}{\sqrt{2}}(\ket{01}+\ket{10}),\\
\ket{\psi^-}=&\frac{1}{\sqrt{2}}(\ket{01}-\ket{10}),\\
\ket{\phi^-}=&\frac{1}{\sqrt{2}}(\ket{00}+\ket{11}),\\
\text{and }\ket{\phi^-}=&\frac{1}{\sqrt{2}}(\ket{00}-\ket{11}),
\end{align}
i.e. they are identical up to local basis changes. The state of Eq.~(\ref{eq:bipartitestate}) is stabilized by the two operators 
\begin{equation}
 g_A= X_A Z_B \text{ and } g_B=Z_A X_B.
\end{equation}
Here
\begin{equation}
 X=\left(
\begin{array}{cc}
 0 & 1 \\
 1 & 0 \\
\end{array}
\right) \text{ and } Z=\left(
\begin{array}{cc}
 1 & 0 \\
 0 & -1 \\
\end{array}
\right)
\end{equation}
are Pauli matrices and the index denotes the party on which this operator acts.\\
Since $g_A$ and $g_B$ are independent, they uniquely define the state of Eq.~(\ref{eq:bipartitestate}). This implies that a repeater scheme is successful if the state produced by the repeater is stabilized by these operators.
\subsection{Transformation of the stabilizer in a circuit}
\label{sec:transformationofstabilizerincircuit}
\begin{figure}[htbp]
 \begin{center}
  \subfigure[$C_Z$\hbox{-}gate]{\hspace{1cm}\includegraphics[scale=0.5]{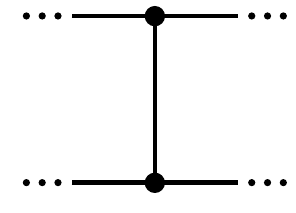}\hspace{1cm}}
  \subfigure[$C_X$\hbox{-}gate]{\hspace{1cm}\includegraphics[scale=0.5]{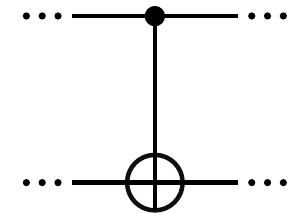}\hspace{1cm}}
 \end{center}
 \caption{The circuit diagram symbols of common entangling gates.}\label{fig:CZ}
\end{figure}

Suppose that $s$ stabilizes the state $\ket{\psi}$ of a system on which now a (unitary) gate $U$ acts. Then 
\begin{equation}
 U s U^\dagger U\ket{\psi} = U \ket{\psi},
\end{equation}
i.e. the operator $U s U^\dagger$ stabilizes the state after the operation of the gate. Because the stabilizer generators uniquely define the quantum state, keeping track of these operators during a quantum computation is equivalent to keeping track of the quantum state. And while it might seem to be more effort to write down and manipulate the set of stabilizer generators than two hold a single quantum state it can be much easier in special situations~\cite{Nielsen00}. This is because the state space increases exponentially with the number of qubits. On the contrary the number of generators increases linearly with the number of qubits and the set of operators occurring as stabilizer generators can be very limited depending on the performed gates. For example it can be restricted to the Clifford group (see Gottesman-Knill-Theorem~\cite{Gottesman97}). This will be the case for quantum repeaters.\\
We need the controlled-Phase gate 
\begin{equation}
 C_Z^{(i,j)}=\ket{0}_i\bra{0}_i \otimes \1_j + \ket{1}_i\bra{1}_i \otimes Z_j \label{eq:CZ}
\end{equation}
that changes the phase of the second qubit if the first qubit is in state $\ket{1}$. Fig.~\ref{fig:CZ} shows the circuit diagram symbol for a $C_Z$ gate.
$C_X$ gates are defined analogously to Eq.~(\ref{eq:CZ}). These are equivalent to $C_Z$ gates up to a local basis change on the second qubit.\\
The $C_Z$ gate transforms the stabilizers $X_A\1_B$ and $\1_A X_B$ that correspond to the product state $\ket{+}_A\ket{+}_B$ into the stabilizers $g_A$ and $g_B$. Thus it is an entangling gate.
\subsection{Inserting and removing intermediate qubits}
\label{sec:insertingandremovingqubits}
\begin{figure}[htp]%
 \begin{center}%
  \subfigure[Application of $C_Z$ gates in two time-steps.]{\includegraphics[width=0.49\linewidth]{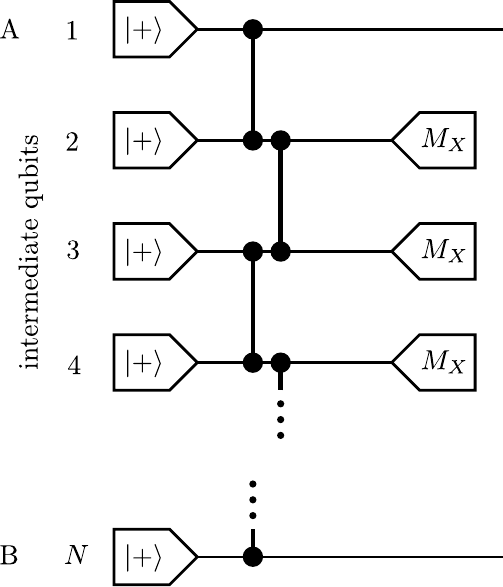}}\hfill%
  \subfigure[Application of $C_Z$ gates in $N-1$ time-steps.]{\includegraphics[width=0.49\linewidth]{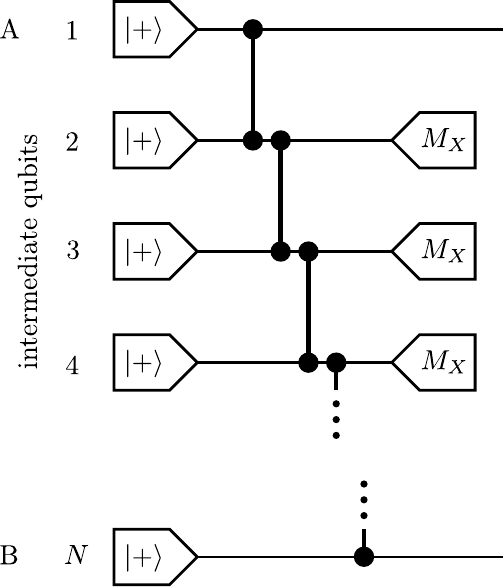}}%
 \end{center}%
 \caption{The basic circuit of quantum repeaters. Intermediate qubits are inserted and measured such that the state of A and B is projected onto a maximally entangled state. The two shown circuits are equivalent, because $C_Z$ gates commute.}\label{fig:basiccircuit}%
\end{figure}

The idea of quantum repeaters is to counter the exponential losses in a fiber by cutting the long transmission line of length $L$ into smaller parts of length $L_0$. The repeater stations connect these shorter channels. Intermediate qubits are inserted, entangled with their neighbors and measured. During this process some kind of error correction (using two-way or one-way communication) is performed. We now describe the basic scheme in the stabilizer formalism. The error correction will be discussed in Section~\ref{sec:logicalerrorrates}. We also ignore the channel for now, since it is not important for understanding why the circuit produces a maximally entangled state. It is included in Section~\ref{sec:repeatercircuit}.\\
Suppose the total number of qubits $N$ is even for simplicity. We sequentially number the qubits from A to B by $1$ to $N$, where $1$ is the qubit of A and $N$ is the one of B. We start with all qubits in the $\ket{+}$ state, i.e. the natural choice of stabilizer generators is $g_i=X_i$. Now neighboring qubits are entangled by $C_Z$ gates. After application of the $C_Z$ gates the list of stabilizers reads
\begin{equation}
\begin{aligned}
g_1=&X_1 Z_2,\\
g_N=&X_N Z_{N-1},\\
\text{and }g_i=&Z_{i-1} X_i Z_{i+1}\text{ for } 1<i<N.
\end{aligned}\label{eq:linegenerators}
\end{equation}
By multiplication of $g_i$ with even and odd $i$ we see that
\begin{align}
 S_A=&X_1 X_3 X_5 ... X_{N-1} Z_N\\
\text{and } S_B=&Z_1 X_2 X_4 X_6 ... X_N
\end{align}
are two stabilizers of the state, respectively. We call these two stabilizers connecting A and B the main stabilizers, because they will play a central role in understanding the quantum repeater in the stabilizer formalism. In the multipartite case discussed at the end of this article, there will be one main stabilizer per party.\\
Now the intermediate qubits ($2,3,...,N-1$) are measured in $X$ basis, because this transforms the stabilizer in the desired way. We replace the corresponding operator in the stabilizer (see Eq.~(\ref{eq:linegenerators})) by the measurement outcome to obtain a stabilizer of the reduced state. The state of A and B after all measurements is stabilized by $\pi_A g_A=\pi_A X_A Z_B$ and $\pi_B g_B=\pi_B Z_A X_B$, where $\pi_A=\pm 1$ and $\pi_B=\pm 1$ are the parities of the measurement outcomes on odd and even qubits, respectively. One can correct for these measurement outcome dependent factors by applying so called by-product operators, here $X_A^{\pi_B} Z_A^{\pi_A}$. After this correction the state is stabilized by $g_A$ and $g_B$ as desired. Therefore the circuit that initializes all qubits in the $\ket{+}$ state, connects all neighboring qubits via $C_Z$ gates and measures the intermediate qubits in the $X$-basis can be used to create a maximally entangled pair shared by A and B.\\
In the context of the original quantum repeater~\cite{Briegel98,Abruzzo13}, the described procedure of projecting onto a bipartite entangled state is usually called entanglement swapping and the application of the by-product operators is called a correction of the ``Pauli-frame''.\\
Note that all the $C_Z$ gates commute. Hence the order of these gates is irrelevant in the ideal case, but it becomes relevant when the error propagation is analyzed. There are mainly two orderings of the gates: sequentially and in two steps (e.g. first gates with odd labeled control qubits, then gates with even labelled control qubits). The two corresponding circuits are shown in Fig.~\ref{fig:basiccircuit}. Apart from the error correction method quantum repeater schemes can also differ in the order of the gates and the position of the transmission channels inside the circuit.\\
The produced state given in Eq.~\ref{eq:bipartitestate} and the state stabilized by the generators given in Eq.~(\ref{eq:linegenerators}) are examples of graph states~\cite{Briegel01,Schlingemann01}. It is possible to create and distribute every graph state in a similar way~\cite{Epping15}. We describe this generalization in Section~\ref{sec:multipartite}.
\section{The error model of depolarizing noise}
\label{sec:errormodel}
As a simple noise model we employ the depolarizing channel $\varepsilon_f(\rho)$~\cite{Nielsen00}. It depends on a parameter $f$ which defines
the strength of the noise. With probability $f$ (for ``failure'') the state $\rho$ is replaced by the completely mixed state $\1/d$, while it remains
$\rho$ with probability $(1-f)$, which leads to the state
\begin{equation}
 \varepsilon_f(\rho)=(1-f)\rho+f \frac{\1}{d}, \label{eq:depolarizingchannel}
\end{equation}
where $d$ is the dimension of the Hilbert space.
\subsection{Error discretization}
\label{sec:errordiscretization}
One can replace the identity term in Eq.~(\ref{eq:depolarizingchannel}) in the single qubit version using that
\begin{equation}
 \frac{\1}{2} = \frac{\rho+X\rho X+ XZ\rho ZX + Z\rho Z}{4}. \label{eq:ID2}
\end{equation}
This leads to the form
\begin{equation}
 \varepsilon_f(\rho)=(1-f)\rho+f\frac{\rho+ X\rho X+XZ\rho ZX+Z\rho Z}{4},
\end{equation}
which has the following interpretation. With probability $(1-f)$ the channel acts as the ideal identity channel, while it ``fails'' with
probability $f$. In case of failure there is a chance of $\frac{1}{2}$ for an $X$ error to occur and an independent chance of $\frac{1}{2}$
for a $Z$ error to occur.\\
The analogous relation to Eq.~(\ref{eq:ID2}) for $n$ qubits reads
\begin{equation}
 \frac{\1}{2^n}=\frac{1}{4^n}\sum_{i_1,i_2,...,i_n\in\{\1,X,XZ,Z\}}  \bigotimes_{k=1}^n i_k\rho \bigotimes_{k=1}^n i_k^\dagger
\end{equation}
and leads to the same error probabilities. We use this relation for the error discretization of $n$-qubit gates and in particular for the $C_Z$ gate. Thus the description of continuous
noise has been replaced by a description in terms of randomly occurring discrete errors $X$ and $Z$~\cite{Nielsen00}. This formulation is convenient with
respect to error propagation and the stabilizer formalism. 
\subsection{Description of erasure errors}
\label{sec:erasures}
We model erasures with the same error model, but in contrast to noise they are noticed in the sense that it is known which qubit is affected. We think of this as a third measurement outcome, a no-detection outcome, that we denote by a ``?''. Analogously to the unnoticed errors, the state of an erased qubit is replaced by the completely mixed state $\1/2$, i.e. it leads to $X$ and $Z$ errors from the viewpoint of discretized errors. The response of the elements of the circuit to erased qubits might strongly depend on the physical implementation and the error model can be improved for specific examples. Notice that this simple model possesses the main property in the context of entanglement distribution: If a qubit gets lost, then it cannot become correlated with any other qubit via a gate that processes them after the loss happened.
\subsection{Error propagation by gates}
\label{sec:errorpropagation}
Gates propagate errors, i.e. errors $e$ before a gate are equivalent to possibly different errors $e'$ after the gate~\cite{Nielsen00}.
Consider an arbitrary gate $U$. An error $e$ before $U$ corresponds to the overall action of $Ue$ onto the state. Due to unitarity of $U$ we
can write \begin{equation}
Ue=\underbrace{UeU^\dagger}_{e'} U =e' U,
\end{equation}
i.e. the error is propagated to an $e'=Ue U^\dagger$ error. Table~\ref{tab:errorpropagation}
lists this relation for the most common cases.\\
Tracking the propagation of $X$ and $Z$ errors in a quantum circuit is a crucial part of the error analysis.
\begin{table}[tp]
 \caption{Propagation of $X$ and $Z$ errors by the  Hadamard- ($H$), controlled-Not- ($C_X$), and controlled-Phase-gate
($C_Z$). Here the index $i$ with $i=1,2$ refers to the $i$-th qubit. Note that you can also read the stabilizer transformation $U s U^\dagger$ from this table.}%
 \label{tab:errorpropagation}%
 \setlength{\tabcolsep}{0em}%
 \begin{center}\begin{tabular}{r@{\,}l}
  $Ue=$ & $e'U$\\
\hline
$H X=$& $Z H$\\
$H Z=$ &$X H$\\
$C_X X_1=$& $X_1 X_2 C_X$\\
$C_X Z_1=$ &$Z_1 C_X$\\
$C_X X_2=$ &$X_2 C_X$\\
$C_X Z_2=$ &$Z_1 Z_2 C_X$\\
$C_Z X_1=$ &$X_1 Z_2 C_Z$\\
$C_Z Z_1=$ &$Z_1 C_Z$\\
$C_Z X_2=$ &$Z_1 X_2 C_Z$\\
$C_Z Z_2=$ &$Z_2 C_Z$
 \end{tabular}\end{center}
\end{table}
\section{Physical errors in quantum repeater circuits}
\label{sec:repeatercircuit}
\begin{figure}[htp]
 \begin{center}
  \includegraphics[width=\linewidth]{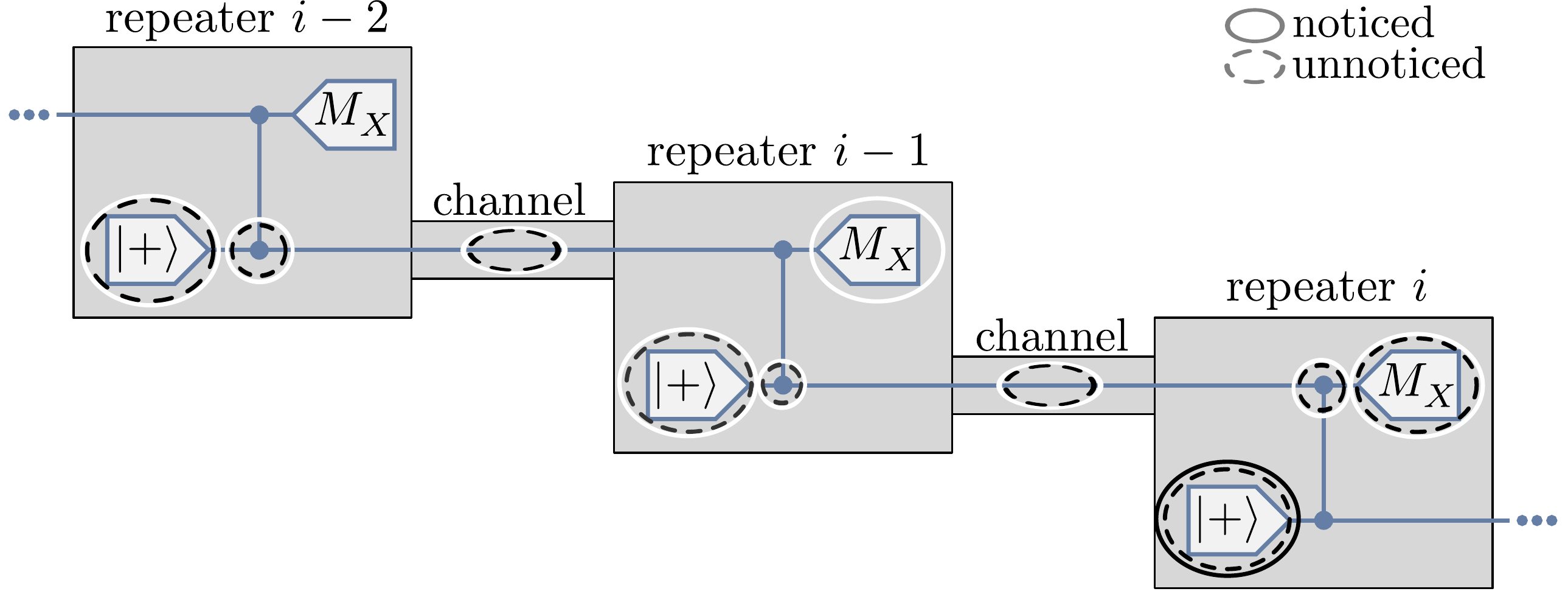}
 \end{center}
 \caption{A basic circuit for a quantum repeater with encoding. Errors at the encircled locations inside the circuit indicate possible causes for a flipped measurement outcome at repeater $i$, see text.  Errors inside white (black) circles contribute to the noticed (unnoticed) error rate at station with number $i$. Solid (dashed) circles denote errors that are noticed (unnoticed).}
 \label{fig:circuit}
\end{figure}

We first analyze the physical error rates of the circuit with a single qubit per station shown in Fig.~\ref{fig:circuit}, which, in contrast to Fig.~\ref{fig:basiccircuit}, now includes the transmission channels. Section~\ref{sec:othercircuits} treats an important variation of this circuit and the error correction is discussed in Section~\ref{sec:logicalerrorrates}.\\
Any operation inside a circuit can cause an error. We use different indices to the symbol $f$ to denote the failure rates of the
corresponding process. These are preparation ($f_P$), transmission ($f_T$), gates ($f_G$) and measurement ($f_M$). We add another index
$u$ or $n$ for unnoticed and noticed errors, respectively. Errors might be noticed by a non-detection event, i.e. no click in some time bin
where we expected one. This gives the additional knowledge of the qubit on which this error occurred. Apart from that we treat these errors using the same model which we introduced in Section~\ref{sec:errormodel}.\\
Typical transmission losses have the form~\cite{Gisin02}
\begin{equation}
f_{T,n}=1-(1-f_{C,n})e^{-\frac{L_0}{L_{\mathrm{att}}}}, \label{eq:fTn}
\end{equation}
where $f_{C,n}$ describes coupling losses, $L_0=\frac{L}{N-1}$ is the repeater spacing and $L_{\mathrm{att}}\approx 20\;\mathrm{km}$ gives the fiber attenuation.\\
To estimate the effective error rate of the measurement outcomes, we collect all
sources of errors that affect the outcome of a specific measurement. An error on the measurement outcome remains, if an odd number of errors
propagate from the source processes to the measurement, while an even number of errors cancels each other. We therefore introduce the
functions
\begin{align}
 \Peven(P,N)=& \frac{1}{2} \left(1 + (1-2 P)^N\right)\\
 \text{and }\Podd(P,N)=& \frac{1}{2} \left(1 - (1-2P)^N\right), \label{eq:PoddPN}
\end{align}
which denote the probability to have an even and odd number of events, respectively, in a sequence of $N$ runs, where in each run the
probability of the event is $P$. We generalize these formulas to the case where the probability of the event differs in each run. These
probabilities are pooled into a vector $\vec{p}$ of dimension $N$ and one can write
\begin{align}
 \Peven\left(\vec{p}\right) =& \sum_{\substack{n=0\\|n|_H\text{ even}}}^{2^N-1} \prod_{k=1}^N p_k^{n^{(k)}}
(1-p_k)^{1-n^{(k)}}\\
 \text{and }\Podd\left(\vec{p}\right) =& \sum_{\substack{n=0\\|n|_H\text{ odd}}}^{2^N-1} \prod_{k=1}^N p_k^{n^{(k)}}
(1-p_k)^{1-n^{(k)}}. \label{eq:Poddp}
\end{align}
Here $|n|_H$ is the Hamming weight of $n$ in binary representation and $n^{(k)}$ is the $k$-th binary digit of $n$. These definitions allow
a compact notation for the exact error rate.\\
Consider the measurement on the repeater station $i$ in Fig.~\ref{fig:circuit}. In total there are ten sources of errors for this
measurement (circles in Fig.~\ref{fig:circuit}): three preparations, three gates, two channels and two measurements. Errors at positions in the circuit other than the shown ones cannot
propagate to the measurement under consideration. We first focus on sources of an unnoticed error of the $X$-measurement on station $i$ (black circles in Fig.~\ref{fig:circuit}).
The measurement outcome is flipped by $Z$ errors, as $Z\ket{+}=\ket{-}$ and $Z\ket{-}=\ket{+}$, but not by $X$ errors as $X\ket{+}=\ket{+}$
and $X\ket{-}=-\ket{-}$. We give a complete list of error causes of an unnoticed error in Table~\ref{tab:unnoticed}.
\begin{table}[tp]
 \caption{All causes of unnoticed measurement errors at repeater $i$. The corresponding processes are marked by a black circle in Fig.~\ref{fig:circuit}.}\label{tab:unnoticed}
 \begin{center}
  \begin{tabular}{c|c|l}
   probability & operator & site \\
\hline
   $\frac{f_{P,u}}{2}$ & $X$ & Preparation of $i-2$ \\
   $\frac{f_{G,u}}{2}$ & $X$ & Gate of $i-2$ \\
   $\frac{f_{T,u}}{2}$ & $X$ & Channel from $i-2$ to $i-1$ \\
   $\frac{f_{P,u}}{2}$ & $Z$ & Preparation of $i-1$ \\
   $\frac{f_{G,u}}{2}$ & $Z$ & Gate of $i-1$ \\
   $\frac{f_{T,u}}{2}$ & $Z$ & Channel from $i-1$ to $i$ \\
   $\frac{f_{G,u}}{2}$ & $Z$ & Gate of $i$ \\
   $\frac{f_{M,u}}{2}$ & $Z$ & Measurement of $i$ \\
   $\frac{f_{P,u}+f_{P,n}}{2}$ & $X$ & Preparation of $i$ \\
   $\frac{f_{G,u}}{2}$ & $X$ & Gate of $i$ \\
  \end{tabular}
 \end{center}
\end{table}
We exemplify the route of an error for the $X$-error occurring with probability $\frac{f_{P,u}}{2}$ in the preparation at $i-2$. It passes
the gate of that station and the subsequent channel. At the repeater $i-1$ it propagates to an $Z$ error on the qubit $i$, passes channel
and gate and flips the measurement outcome.\\
Noticed errors on the qubit that is measured at station $i-1$ have a high probability of
50\,\% to lead to a flipped measurement outcome at site $i$. We thus choose to mark the outcome of that measurement as ``?''. In this way we
exclude these noticed errors from the unnoticed error rate of $i$, which reads
\begin{equation}
\begin{aligned}
f_u=&\Podd
\left(\left(\Podd\left(\frac{f_{P,u}}{2},2\right),\right. \frac{f_{P,n}+f_{P,u}}{2},\right.\\
    &\Podd\left.\left.\left(\frac{f_{G,u}}{2},3\right),\Podd\left(\frac{f_{T,u}}{2},2\right),\frac{f_{M,u}}{2}
\right)\right).\\ 
\end{aligned} \label{eq:fu}
\end{equation}
The full list of sources of noticed errors in the measurement at site $i$ is given by Table~\ref{tab:noticed} and in Fig.~\ref{fig:circuit} white circles mark the corresponding positions in the circuit.
\begin{table}[tp]
 \caption{All causes of noticed measurement errors at repeater $i$. The corresponding processes are marked by a white circle in Fig.~\ref{fig:circuit}.}\label{tab:noticed}
 \begin{center}
  \begin{tabular}{c|c|l}
   probability & operator & site \\
\hline
   $\frac{f_{P,n}}{2}$ & $X$ & Preparation of $i-2$ \\
   $\frac{f_{G,n}}{2}$ & $X$ & Gate of $i-2$ \\
   $\frac{f_{T,n}}{2}$ & $X$ & Channel from $i-2$ to $i-1$ \\
   $\frac{f_{M,n}}{2}$ & $Z$ & Measurement of $i-1$ \\
   $\frac{f_{P,n}}{2}$ & $Z$ & Preparation of $i-1$ \\
   $\frac{f_{G,n}}{2}$ & $Z$ & Gate of $i-1$ \\
   $\frac{f_{T,n}}{2}$ & $Z$ & Channel from $i-1$ to $i$ \\
   $\frac{f_{G,n}}{2}$ & $Z$ & Gate of $i$ \\
   $\frac{f_{M,n}}{2}$ & $Z$ & Measurement of $i$ \\
  \end{tabular}
 \end{center}
\end{table}
The outcome is ``?'' if \emph{any} of these errors occurred. This happens with probability
\begin{equation}
f_n = 1-(1-f_{P,n})^2(1-f_{G,n})^3(1-f_{T,n})^2(1-f_{M,n})^2. \label{eq:fn}
\end{equation}
\subsection{How far do errors propagate?}
\label{sec:howfar}
On the first glance it might seem possible that errors propagate along the whole line of repeater stations to Bob. This is not the case. The
measurement outcome on repeater $i$ is only affected by errors on repeater stations $i-2$ to $i$. The $C_Z$ gates propagate $X$ to $Z$
errors on the neighboring qubit. These do not propagate across $C_Z$ gates. Thus only elements of the circuit that involve a neighboring
qubit of the one measured in qubit $i$ need to be considered. For a full error analysis all these sources need to be included. In
particular, it is usually not exhaustive to consider only a single repeater station independently of the previous ones. One has to pay
attention to such restrictions when comparing different repeater schemes from the literature.
\subsection{Bit flip errors caused by erasures}
\label{sec:errorsbylosses}
The effect of one lost qubit before the application of a two-qubit gate may strongly depend on the physical implementation of the gate. It
is reasonable, however, to assume that there will be some unwanted effect on the remaining second qubit. In our error model the lost qubit
is replaced by the completely mixed state, or equivalently, $X$ and $Z$ errors randomly occur at the position of the loss. These errors
propagate across the two-qubit gates, possibly leading to flipped outcomes of measurements on these adjacent qubits. In this way losses in our
model lead to noise on detected qubits.
\subsection{Other circuits}
\label{sec:othercircuits}
\begin{figure}[htp]
% \begin{subfigure}[t]{0.5\textwidth}
  \centering
  \subfigure[Spatial diagram]{\includegraphics[scale=0.5]{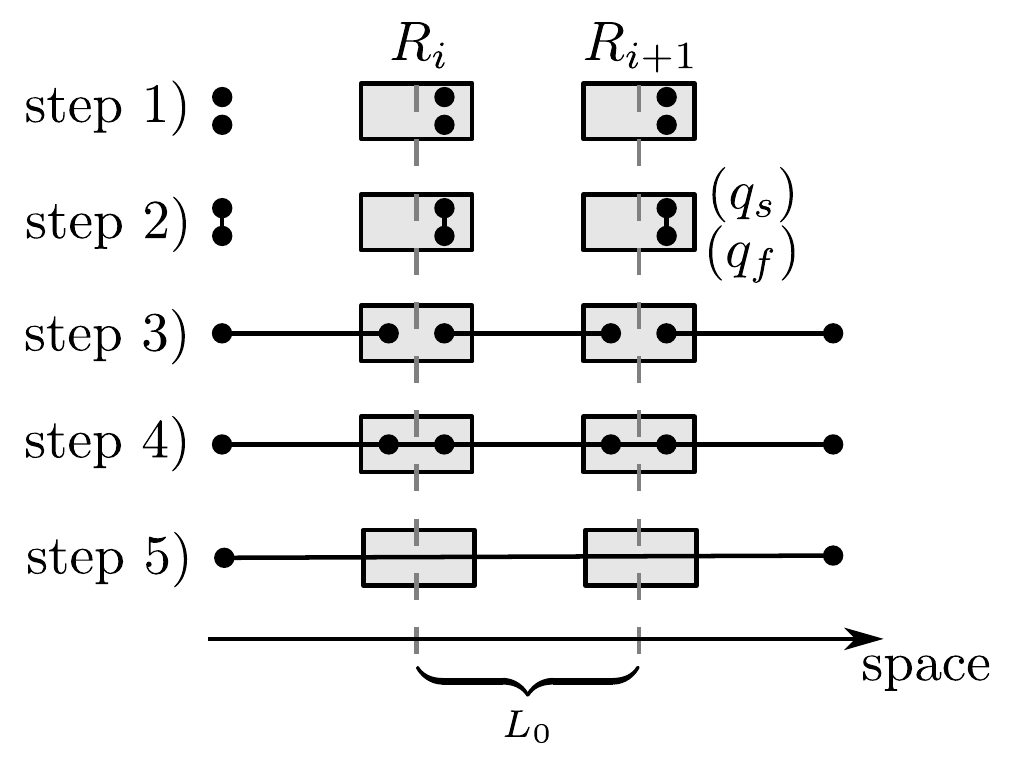}}
  \subfigure[Circuit diagram]{\includegraphics[scale=0.5]{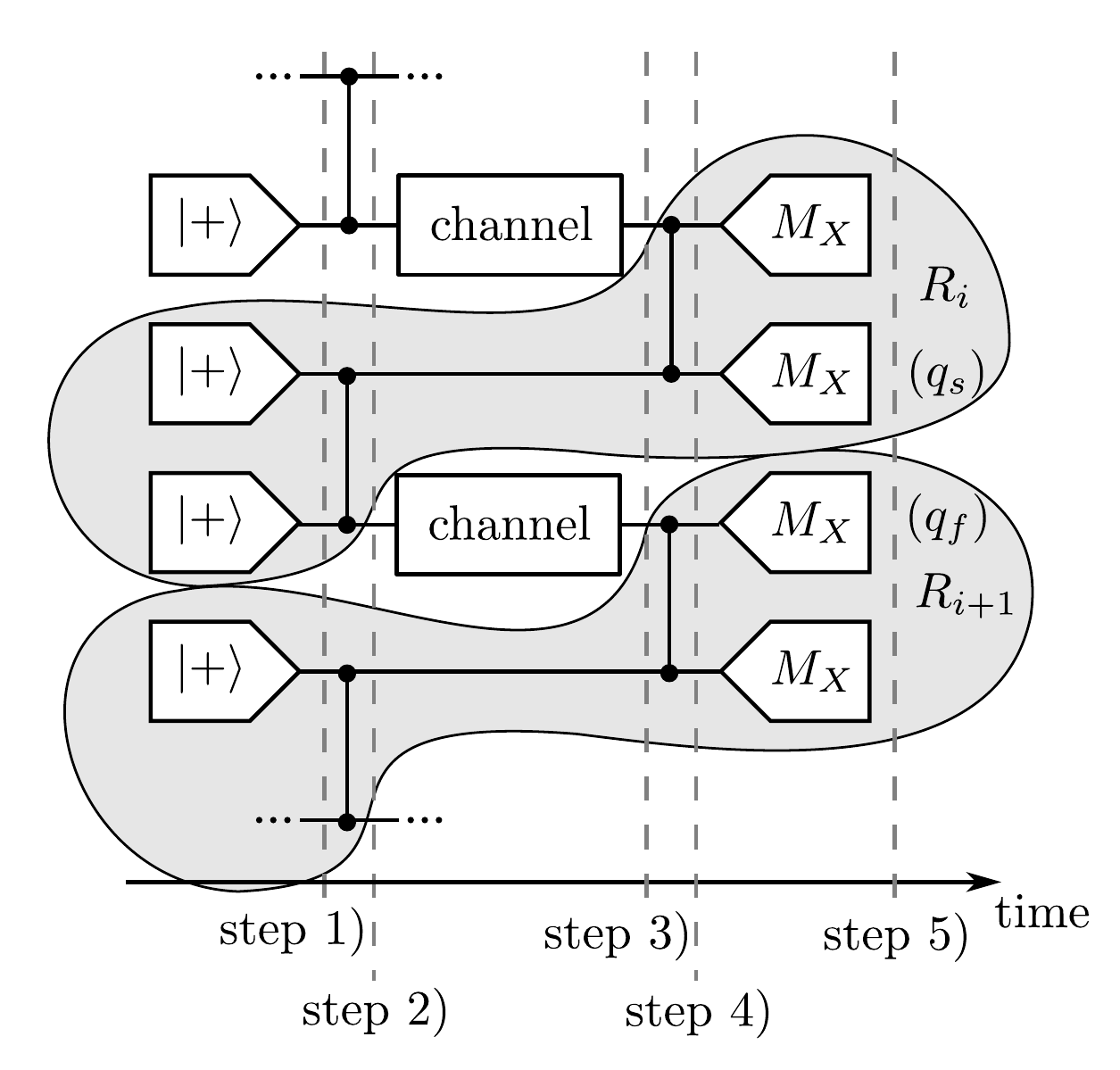}}
% \end{subfigure}
 \caption{Schematic of a repeater with two qubits per station. Two repeaters $R_i$ and $R_{i+1}$ are shown in a spatial diagram (a) and the circuit diagram (b). 
Entangled pairs are created, distributed amongst neighboring repeater stations and then connected locally. In a last step local measurements project onto a two-qubit entangled state.}
\label{fig:connectingpairs}
\end{figure}

An analogous error analysis can be done for other circuits, too. Here we discuss the error propagation in circuits where only
half of the qubits are transmitted through the channel, while the other half remains stationary as another
example. Fig.~\ref{fig:connectingpairs} shows a schematic of such a repeater and the corresponding
circuit.
Again we identify all sources of an flipped measurement outcome. The treatment of noticed errors differs for stationary and flying
qubits, so we calculate two different error rates. The index $s$ or $f$ denotes stationary or flying qubits, respectively.
The rates of unnoticed errors read
\begin{align}
 \begin{split}f_{q,s}=&
\Podd\left(\left(\frac{f_{P,u,f}}{2},\frac{f_{G,u,f}}{2},\frac{f_{T,u,f}}{2},\frac{f_{P,u,s}}{2},\frac{f_{G,u,s}}{2},\right.\right.\\
   & \left.\left.\frac{f_{G,u,s}}{2},\frac{f_{M,u,s}}{2},\frac{f_{P,u,f}+f_{P,n,f}}{2}
\right)\right),
\end{split}\\
\begin{split}f_{q,f}=&\Podd
\left(\left( \frac{f_{P,u,s}+f_{P,n,s}}{2},\frac{f_{P,u,f}}{2},\frac{f_{T,u,f}}{2},\frac{f_{G,u,f}}{2},\right.\right.\\
   & \left.\left. \frac{f_{M,u,f}}{2},\frac{f_{P,u,s}+f_{P,n,s}}{2},\frac{f_{G,u,s}+f_{G,n,s}}{2}
\right)\right)
\end{split}
\end{align}
and the rates of noticed errors read
\begin{align}
 \begin{split}
f_{l,s}=& 1-(1-f_{P,n,f})(1-f_{G,n,f})(1-f_{T,n,f})\times\\
&\times(1-f_{M,n,f})(1-f_{P,n,s})(1-f_{G,n,s})(1-f_{M,n,s})
 \end{split}\\
\begin{split}
 f_{l,f}=&1-(1-f_{P,n,f})(1-f_{G,n,f})^2(1-f_{T,n,f})\times\\
&(1-f_{M,n,f}).
\end{split}
\end{align}
Note that, analogously to the other circuit, we choose to mark the stationary qubit as lost whenever the previous flying qubit got lost.
This is not necessary but improves the error correction, as the stationary qubit has a high probability for errors in this case.
\section{Logical error rates of encoded qubits}
\label{sec:logicalerrorrates}
So far we considered the error rates on physical qubits. Quantum repeater with encoding use error correction codes~\cite{Shor95,Steane96a,Bennett96,Knill00} to encode the
information of logical qubits into a larger number of physical qubits. The circuits discussed above are shifted to the logical level, i.e.
the shown qubits and operations are now replaced by their logical counterparts. Before going into the details of the analysis of the logical
errors we give a short reminder of Calderbank-Shor-Steane (CSS) codes~\cite{Calderbank96,Steane96}.
\subsection{Calderbank-Shor-Steane codes}
\label{sec:steanecode}
\begin{figure}[tp]
\begin{center}
\includegraphics[scale=0.5]{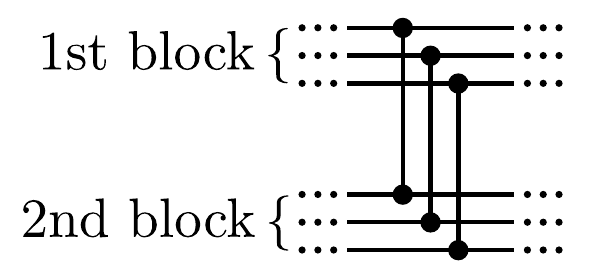}
\end{center}
\caption{Transversal implementation of a $C_Z$ gate. The $i$-th gate acts on the $i$-th physical qubits of the first and second block.}\label{fig:transversal}
\end{figure}

Stabilizer codes can be defined via the generators of the stabilizer of the code space $g_1,...,g_{n-k}$ ($n$ and $k$ are the numbers of
physical and logical qubits, respectively)~\cite{Gottesman96}. Valid codewords $\ket{\psi}$ satisfy $g_i \ket{\psi}=\ket{\psi}$. The logical $\bar{Z}_i$
operators, $i=1,2,...,k$, are chosen such that they commute with and are independent from each other and the stabilizer generators. If the last are tensor products of either only $X$ and $\1$ or only $Z$ and $\1$, the code is called a CSS code. We give the popular example of the Seven-Qubit-Steane code in Table~\ref{tab:steanecode}.\\
\begin{table}[htp]
\caption{The stabilizer generators and logical operators of the Seven-Qubit-Steane code.}\label{tab:steanecode}
\begin{center}
\setlength{\tabcolsep}{1pt}
\begin{tabular}{ccccccccccccccc}
$g_1$&$=$ & $\1$&$\otimes $&$\1$&$\otimes$&$\1$&$\otimes$&$ X$&$\otimes$&$ X$&$ \otimes$&$ X$&$ \otimes $&$X$\\
$g_2$&$=$ & $\1$&$\otimes$&$ X$&$\otimes$&$ X$&$\otimes$&$ \1$&$\otimes $&$\1 $&$\otimes $&$X$&$ \otimes $&$X$\\
$g_3$&$=$ & $X$&$\otimes$&$ \1$&$\otimes$&$ X$&$\otimes$&$ \1$&$\otimes $&$X$&$ \otimes $&$\1 $&$\otimes$&$ X$\\
$g_4$&$=$ & $\1$&$\otimes$&$ \1$&$\otimes$&$\1$&$\otimes$&$ Z$&$\otimes$&$ Z$&$ \otimes$&$ Z$&$ \otimes$&$ Z$\\
$g_5$&$=$ & $\1$&$\otimes$&$ Z$&$\otimes$&$ Z$&$\otimes$&$ \1$&$\otimes$&$ \1$&$ \otimes $&$Z$&$ \otimes $&$Z$\\
$g_6$&$=$ & $Z$&$\otimes$&$\1$&$\otimes$&$ Z$&$\otimes$&$ \1$&$\otimes$&$ Z$&$ \otimes $&$\1$&$\otimes$&$ Z$\\[1ex]

$\bar{Z}$&$=$ & $Z$&$\otimes$&$ Z$&$\otimes$&$ Z$&$\otimes$&$ Z$&$\otimes$&$ Z$&$ \otimes$&$ Z$&$ \otimes$&$ Z$\\
$\bar{X}$&$=$ & $X$&$\otimes$&$ X$&$\otimes$&$ X$&$\otimes$&$ X$&$\otimes$&$ X$&$ \otimes$&$ X$&$ \otimes$&$ X$
\end{tabular}
\end{center}
\end{table}
The transversal, i.e. qubitwise (see Fig.~\ref{fig:transversal}), application of controlled-NOT gates $C_X=\ket{0}\bra{0}\otimes \1+\ket{1}\bra{1}\otimes X$ performs the following mapping of the stabilizer generators. If $g_i$ contains only $X$ operators, then $g_i\otimes \1 \rightarrow g_i\otimes g_i$, $\1\otimes g_i\rightarrow \1\otimes g_i$, while a $g_i$ containing $Z$ operators is mapped according to $g_i\otimes \1\rightarrow g_i\otimes \1$ and $\1\otimes g_i\rightarrow g_i\otimes g_i$ (see Table~\ref{tab:errorpropagation}). Thus transversal application of $C_X$ is a valid gate in CSS codes, i.e. it preserves validity of the codeword.\\
The Seven-Qubit-Steane code and the quantum Golay code have even more symmetry: Exchanging $X$ and $Z$ operators in a stabilizer operator $s$ leads to another element of the stabilizer $s'$. This implies that the transversal Hadamard gate is valid and hence also the transversal controlled-Phase gate $C_Z$ (Fig.~\ref{fig:transversal}).\\
Transversal implementations of gates have advantageous error propagation properties: Because a single error on one block cannot lead to more than one error on the other block, these errors remain correctable after the application of the gate. If all gates are implemented transversally, then the physical error rates do not depend on the code or its size. Thus Eqs.~(\ref{eq:fu}) and (\ref{eq:fn}) are true for all CSS codes.
\subsection{Ideal measurement outcomes are codewords}
\label{sec:outcomesarecodewords}
The stabilizer generators that contain $X$ correspond to the rows of the parity-check matrix of a (classical) linear block code. The classical parity-check matrix is obtained from the stabilizer generators by replacing a $\1$ by $0$ and a $X$ by $1$ (and $\otimes$ by whitespace). The parity-check matrix $H$ of a classical code can be used to check whether some word $\vec{c}$ is inside the code space, because $H\vec{c}^T=\vec{0}$ if and only if $\vec{c}$ is a codeword. In absence of any errors, the measurement of any stabilizer generator containing $X$ operators gives a $+1$ result and, equivalently, the vector $\vec{c}$ of the individual $X$-measurements passes the parity check. That is, this vector of the $X$-measurement outcomes is a codeword of the associated classical linear block code. 
\subsection{Calculating the logical error rate}
\label{sec:logicalerrorrate}
After the $\bar{X}$-measurement we are dealing with classical data. In the presence of imperfections, some of the bits will be flipped. Some values are marked as ``?'' due to a non-detection event. This data could have been generated by a classical channel with both bit flip and erasure errors. Thus a classical decoder can be used to find and correct the errors on the data.\\
Some loss patterns in the data are not likely to be corrected. In this case it can be beneficial to abort the protocol and throw away the data, i.e. rerun the experiment. This leads to a success probability of the protocol. If $\mathcal{F}$ is the set of fatal error patterns on which we choose to abort, then the success probability of the protocol is
\begin{equation}
\Psucc=1-\sum_{e\in\mathcal{F}} P_\mathrm{e}(e),
\end{equation}
where $P_\mathrm{e}(e)$ is the probability of the error pattern $e$. The impact of the choice of $\mathcal{F}$ on the performance of the protocol with respect to some figure of merit is discussed in Section~\ref{sec:abortion}.\\
If the protocol has not been aborted, then after decoding we are left with a valid codeword (but not necessarily the correct one), from which we can calculate the $\bar{X}$ outcome, which is the parity of the bits that contribute to $\bar{X}$ (i.e. the positions where $\bar{X}$ contains a $\1$ are excluded).\\
The logical error rate $\bar{f}_u$ is the probability to arrive at the wrong $\bar{X}$ outcome when following the above procedure. Averaged over the logical qubits of one block we get
\begin{equation}
\bar{f}_u(f_u,f_n)=\frac{1}{k}\sum_{e\not\in \mathcal{F}} f(e) P_\mathrm{e} (e), \label{eq:fbarq}
\end{equation}
where $f(e)$ is the number of wrong logical outcomes in a single block. For small codes this can be easily calculated by trying the decoder on any possible error pattern. For larger codes this calculation cannot be done by ``brute-force'' anymore and more clever approaches are necessary.\\
We explicitly performed the sum in Eq.~(\ref{eq:fbarq}) for the Seven-Qubit-Steane code and the fatal error set
\begin{equation}
\mathcal{F}=\{e|e \text{ contains more than $n_{\max}$ losses}\}
\end{equation}
for $n_{\max}=1,2,...,7$. The results are listed in Table~\ref{tab:steanefubar}.
\begin{table*}[tp]
\caption{Logical error rates of the Steane code for different abortion strategies: $n_{\max}$ is the maximal number of tolerated losses before abortion.}\label{tab:steanefubar}
\setlength{\tabcolsep}{1pt}
\begin{tabular}{rl} 
\multicolumn{2}{l}{$n_{\mathrm{max}}=0$:}\\
$\bar{f}_u(f_u,f_n)=$ &$\left(f_n-1\right)^7 f_u^2 \left(48 f_u^5-168 f_u^4+252 f_u^3-210 f_u^2+98 f_u-21\right)$ \\
$P_{\mathrm{succ}}(f_n)=$ & $\left(1-f_n\right)^7$\\[1ex]
\multicolumn{2}{l}{$n_{\mathrm{max}}=1$:}\\
$\bar{f}_u(f_u,f_n)=$  & $\left(f_n-1\right)^6 f_u \left(48 \left(f_n-1\right) f_u^6-168 \left(f_n-1\right) f_u^5+252 \left(f_n-1\right) f_u^4-210 \left(f_n-1\right) f_u^3\right.$\\
& $\left.+14 \left(9 f_n-7\right) f_u^2+21 \left(1-3 f_n\right) f_u+21 f_n\right)$\\
$P_{\mathrm{succ}}(f_n)=$ & $\left(f_n-1\right)^6 \left(6 f_n+1\right)$ \\[1ex]
\multicolumn{2}{l}{$n_{\mathrm{max}}=2$:}\\
$\bar{f}_u(f_u,f_n)=$  & $\left(f_n-1\right){}^5 f_u \left(48 \left(f_n-1\right){}^2 f_u^6-168 \left(f_n-1\right){}^2 f_u^5+252 \left(f_n-1\right){}^2 f_u^4-210 \left(f_n-1\right){}^2 f_u^3\right.$\\
&$\left.+14 \left(f_n \left(3 f_n-16\right)+7\right) f_u^2+21 \left(f_n \left(3 f_n+4\right)-1\right) f_u-21 f_n \left(2 f_n+1\right)\right)$ \\
$P_{\mathrm{succ}}(f_n)=$ & $-\left(f_n-1\right){}^5 \left(15 f_n^2+5 f_n+1\right)$\\[1ex]
\multicolumn{2}{l}{$n_{\mathrm{max}}=3$:}\\
$\bar{f}_u(f_u,f_n)=$  & $\frac{1}{2} \left(f_n-1\right){}^4 \left(f_n^3 \left(96 f_u^7-336 f_u^6+504 f_u^5-420 f_u^4+308 f_u^3-210 f_u^2+84 f_u+7\right)\right.$\\
&$-2 f_n^2 f_u \left(144 f_u^6-504 f_u^5+756 f_u^4-630 f_u^3+266 f_u^2-21 f_u-21\right)$\\
&$+2 f_n f_u \left(144 f_u^6-504 f_u^5+756 f_u^4-630 f_u^3+322 f_u^2-105 f_u+21\right)$\\
&$\left.+2 f_u^2 \left(-48 f_u^5+168 f_u^4-252 f_u^3+210 f_u^2-98 f_u+21\right)\right)$\\
$P_{\mathrm{succ}}(f_n)=$ & $\left(f_n-1\right){}^4 \left(20 f_n^3+10 f_n^2+4 f_n+1\right)$\\[1ex]
\multicolumn{2}{l}{$n_{\mathrm{max}}=4$:}\\
$\bar{f}_u(f_u,f_n)=$ & $\frac{1}{2} \left(f_n-1\right){}^3 \left(3 f_n^4 \left(32 f_u^7-112 f_u^6+168 f_u^5-140 f_u^4+84 f_u^3-42 f_u^2+14 f_u-7\right)\right.$\\
&$-f_n^3 \left(384 f_u^7-1344 f_u^6+2016 f_u^5-1680 f_u^4+840 f_u^3-252 f_u^2+42 f_u+7\right)$\\
&$+12 f_n^2 f_u^2 \left(48 f_u^5-168 f_u^4+252 f_u^3-210 f_u^2+98 f_u-21\right)$\\
&$-6 f_n f_u \left(64 f_u^6-224 f_u^5+336 f_u^4-280 f_u^3+140 f_u^2-42 f_u+7\right)$\\
&$\left.+2 f_u^2 \left(48 f_u^5-168 f_u^4+252 f_u^3-210 f_u^2+98 f_u-21\right)\right)$\\
$P_{\mathrm{succ}}(f_n)=$ & $-15 f_n^7+35 f_n^6-21 f_n^5+1$\\[1ex]
\multicolumn{2}{l}{$n_{\mathrm{max}}=5$:}\\
$\bar{f}_u(f_u,f_n)=$ & $\frac{1}{2} \left(f_n-1\right){}^2 \left(6 f_n^5 f_u \left(16 f_u^6-56 f_u^5+84 f_u^4-70 f_u^3+42 f_u^2-21 f_u+7\right)\right.$\\
&$-2 f_n^4 \left(240 f_u^7-840 f_u^6+1260 f_u^5-1050 f_u^4+546 f_u^3-189 f_u^2+42 f_u-7\right)$\\
&$+f_n^3 \left(960 f_u^7-3360 f_u^6+5040 f_u^5-4200 f_u^4+2016 f_u^3-504 f_u^2+42 f_u+7\right)$\\
&$-6 f_n^2 f_u \left(160 f_u^6-560 f_u^5+840 f_u^4-700 f_u^3+336 f_u^2-84 f_u+7\right)$\\
&$+2 f_n f_u \left(240 f_u^6-840 f_u^5+1260 f_u^4-1050 f_u^3+518 f_u^2-147 f_u+21\right)$\\
&$\left.+2 f_u^2 \left(-48 f_u^5+168 f_u^4-252 f_u^3+210 f_u^2-98 f_u+21\right)\right)$\\
$P_{\mathrm{succ}}(f_n)=$ & $6 f_n^7-7 f_n^6+1$\\[1ex]
\multicolumn{2}{l}{$n_{\mathrm{max}}=6$:}\\
$\bar{f}_u(f_u,f_n)=$ & $f_n^7 \left(48 f_u^7-168 f_u^6+252 f_u^5-210 f_u^4+126 f_u^3-63 f_u^2+21 f_u-\frac{7}{2}\right)$\\
&$-\frac{21}{2} f_n^6 \left(2 f_u-1\right){}^3 \left(4 f_u^4-8 f_u^3+6 f_u^2-2 f_u+1\right)$\\
&$+\frac{21}{2} f_n^5 \left(2 f_u-1\right){}^3 \left(12 f_u^4-24 f_u^3+18 f_u^2-6 f_u+1\right)$\\
&$-105 f_n^4 f_u \left(2 f_u-1\right){}^3 \left(2 f_u^3-4 f_u^2+3 f_u-1\right)+\frac{7}{2} f_n^3 \left(2 f_u-1\right){}^3 \left(60 f_u^4-120 f_u^3+90 f_u^2-30 f_u-1\right)$\\
&$-63 f_n^2 f_u \left(2 f_u-1\right){}^3 \left(2 f_u^3-4 f_u^2+3 f_u-1\right)+21 f_n f_u \left(2 f_u-1\right){}^3 \left(2 f_u^3-4 f_u^2+3 f_u-1\right)$\\
&$+f_u^2 \left(-48 f_u^5+168 f_u^4-252 f_u^3+210 f_u^2-98 f_u+21\right)$\\
$P_{\mathrm{succ}}(f_n)=$ & $1-f_n^7$\\[1ex]
\multicolumn{2}{l}{$n_{\mathrm{max}}=7$:}\\
$\bar{f}_u(f_u,f_n)=$ & $3 f_n^7 \left(2 f_u-1\right){}^3 \left(2 f_u^4-4 f_u^3+3 f_u^2-f_u+1\right)-\frac{21}{2} f_n^6 \left(2 f_u-1\right){}^3 \left(4 f_u^4-8 f_u^3+6 f_u^2-2 f_u+1\right)$\\
&$+\frac{21}{2} f_n^5 \left(2 f_u-1\right){}^3 \left(12 f_u^4-24 f_u^3+18 f_u^2-6 f_u+1\right)-105 f_n^4 f_u \left(2 f_u-1\right){}^3 \left(2 f_u^3-4 f_u^2+3 f_u-1\right)$\\
&$+\frac{7}{2} f_n^3 \left(2 f_u-1\right){}^3 \left(60 f_u^4-120 f_u^3+90 f_u^2-30 f_u-1\right)-63 f_n^2 f_u \left(2 f_u-1\right){}^3 \left(2 f_u^3-4 f_u^2+3 f_u-1\right)$\\
&$+21 f_n f_u \left(2 f_u-1\right){}^3 \left(2 f_u^3-4 f_u^2+3 f_u-1\right)+f_u^2 \left(-48 f_u^5+168 f_u^4-252 f_u^3+210 f_u^2-98 f_u+21\right)$\\
$P_{\mathrm{succ}}(f_n)=$ & $1$\\
\end{tabular}
\end{table*}
For larger codes, like the Golay code~\cite{WilliamsSloane78}, the logical error rate can be taken from the literature~\cite{Elia95}. There the probability $p_w$ that the decoding outputs the wrong codeword is given. Half of the codewords have even and half have odd parity. We therefore assume that the probability of a logical error is $\bar{f}_u=\frac{p_w}{2}$.

\section{The final state}
\label{sec:finalstate}
We motivated in Section~\ref{sec:thecircuit}, that the described circuits produce a maximally entangled state and how this can be understood in the stabilizer formalism. The same reasoning still holds when the operators are shifted to the logical level. The logical state is stabilized by the logical stabilizers, which transform under the action of logical gates analogously to the physical stabilizers. Remember that the state before the measurements is stabilized by the main stabilizers $S_A$ and $S_B$ and thus after the $\bar{X}$-measurements it is stabilized by $g_A$ and $g_B$ up to byproduct operators. These byproduct operators depend on the measurement outcomes. They are necessary even in the ideal case, where all operations and measurements are perfect.\\
Odd numbers of logical errors on the same main stabilizer lead to the wrong parity and thus to the application of the wrong byproduct operators, which implies that a state orthogonal to the intended state given in Eq.~(\ref{eq:bipartitestate}) is produced. We use the symbols $e_A$ and $e_B$ for the two corresponding error rates on the final state. They read
\begin{align}
 e_A=&\Podd \left(\bar{f}_u, \left\lfloor\frac{N-2}{2}\right\rfloor\right)\\
\text{and } e_B=&\Podd \left(\bar{f}_u, \left\lceil\frac{N-2}{2}\right\rceil\right),
\end{align}
and can be interpreted as $X$- and $Z$-error rates on qubit $1$ of Alice. Thus the fidelity of the state is
\begin{equation}
 F=(1-e_A)(1-e_B).
\end{equation}
\subsection{The secret fraction and the costs}
\label{sec:secretfraction}
A very important application of quantum repeaters is with respect to quantum key distribution. In this case one is not interested in the fidelity of the state but in the number of secret bits one can gain from many copies of the state in a quantum key distribution protocol. The ratio of secret bits per distributed entangled state is called secret fraction and in the standard BB84 protocol it is given by
\begin{equation}
 r_{\infty}=\max\{1-h(e_A)-h(e_B),0\},
\end{equation}
where $h(p)=-p\log_2 (p)-(1-p)\log_2 (1-p)$ is the binary entropy. The secret key rate of a quantum repeater,
\begin{equation}
 R_{\mathrm{QKD}}=R_{\mathrm{raw}} r_{\infty}, \label{eq:R}
\end{equation}
is the product of the raw key generation rate $R_{\mathrm{raw}}$ and the secret fraction $r_{\infty}$. If we set the probability of matching basis choice of Alice and Bob (``sifting'') to $1$, which is possible in the asymptotic case~\cite{something}, then $R_{\mathrm{raw}}$ corresponds to the generation rate of entangled states. In a forward error correction scheme this repetition rate of the repeater is basically given by the fundamental time needed for processing the signal at a single repeater station and the success probability of the protocol. We assume that the speed of the operations is limited by the time $T_M$ needed for the measurement at the repeater station. In this case
\begin{equation}
 R_{\mathrm{raw}} = \frac{\Psucc}{T_M}.
\end{equation}
For simplicity we will set the fundamental time $T_M$ to $1$ when considering forward error correction schemes only.
In an attempt to do a fair comparison between repeater schemes with different codes, we use the cost function
\begin{equation}
 C'=\frac{N n}{R_{\mathrm{QKD}} L}\label{eq:Cprime}
\end{equation}
as a figure of merit~\cite{Muralidharan14}. Here $N$ is the number of encoded blocks, $n$ is the number of physical qubits per block, $R_{\mathrm{QKD}}$ is the secret key rate and $L$ is the total distance bridged by the line of repeater stations.
\subsection{The impact of abortion strategies}
\label{sec:abortion}
\begin{figure}[htp]
 \begin{center}
  \includegraphics[width=\linewidth]{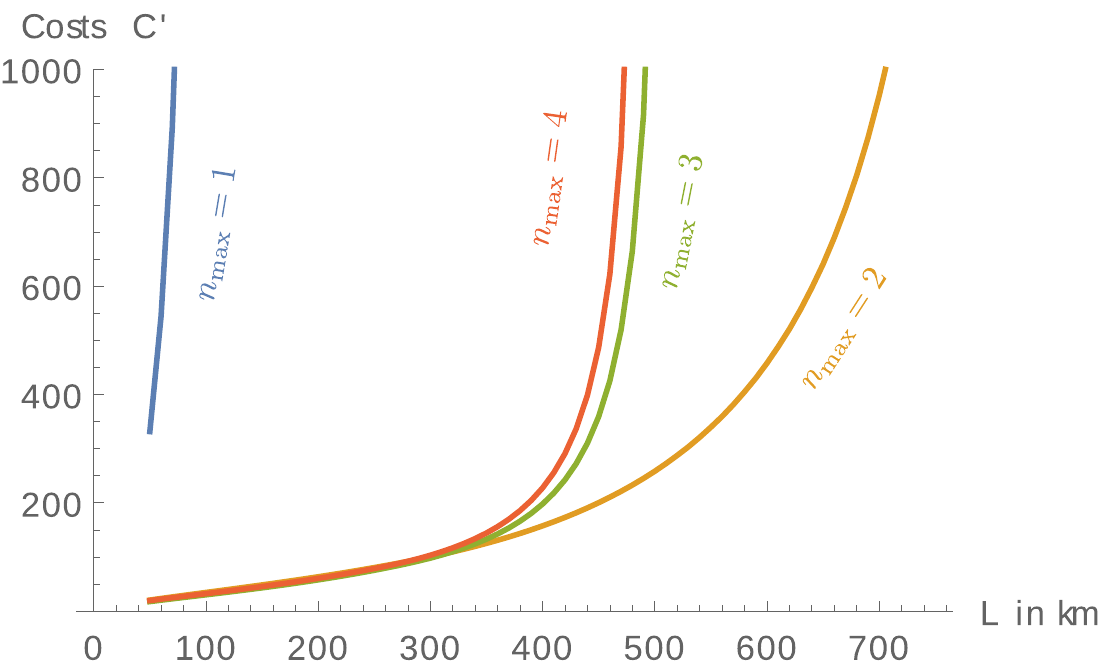}
 \end{center}
 \caption{The cost of the repeater using the Seven-Qubit-Steane code for different abortion strategies. The number $n_{\max}$ denotes the maximal number of tolerated losses. The gate failure rate is $f_G=10^{-4}$ for this plot.}\label{fig:abortionstrategies}
\end{figure}

In the previous sections we derived all the necessary formulas to compare different strategies of encoding. We start the discussion of this result by comparing different abortion strategies $\mathcal{F}$ for the simple Seven-Qubit-Steane code, see Table~\ref{tab:steanecode}. It is based on the (7,4)-Hamming code, which has a Hamming distance of $d=3$. This implies that it can correct $\frac{d-1}{2}=1$ unnoticed errors or $d-1=2$ noticed errors.\\
More noticed errors are unlikely to be corrected and thus an abortion of the protocol will prevent the production of too noisy states. Abortion on two or less losses decreases the success probability unnecessarily. One might therefore expect, that $n_{\max}=2$ gives the optimal fatal error set $\mathcal{F}$. Fig.~\ref{fig:abortionstrategies} supports these considerations.

\subsection{The distillation based protocol}
\label{sec:distillation}
\begin{figure}[htp]%
 \begin{center}%
  \includegraphics[width=\linewidth]{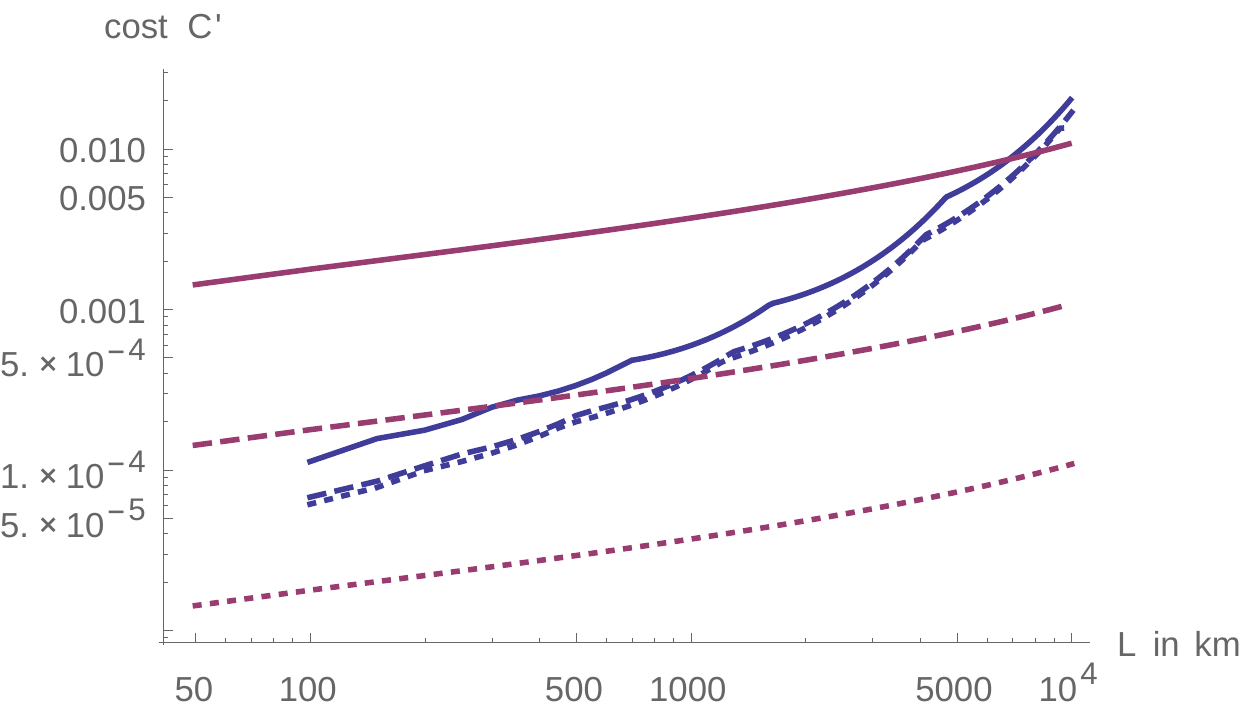}%
 \end{center}%
 \caption{The costs $C'$ (in qubit seconds per bit and kilometer) of the standard repeater (blue) and for the one with Golay code (purple) as a function of the total distance $L$. The gate errors are $f_G=10^{-3}$, other errors are neglected. The measurement time is $T_M=1\;\mathrm{\mu s}$ (dotted), $T_M=10\;\mathrm{\mu s}$ (dashed), and $T_M=100\;\mathrm{\mu s}$ (solid).}\label{fig:originalrepeater}%
\end{figure}

We compare the costs of repeaters with encoding to the standard repeater with two-way communication using the results of~\cite{Abruzzo13}. There the repeater rate is calculated (amongst others) for the following setup. The total distance is divided into $2^{\tilde{N}}$ shorter channels of length $L_0=L/2^{\tilde{N}}$ by repeater stations. Initially $2^{\tilde{k}}$, $\tilde{k}\in\N\cup\{0\}$ entangled states of fidelity $F_0$ w.r.t. some maximally entangled state are distributed amongst neighboring repeater stations. Here two-way classical communication is necessary in order to acknowledge success of the distribution. After $k$ rounds of distillation using the protocol of \cite{Deutsch96} for each channel a single pair with higher fidelity is left (if the initial fidelity is greater than $\frac{1}{2}$). Afterwards a Bell measurement on each repeater station projects onto the final entangled state shared by Alice and Bob.\\
The rate $R^O_{QKD}$ is (to some extent) limited by the classical communication time which is necessary to acknowledge the successful transmission and distillation.
For the two-way protocol we incorporate the measurement time $T_M$ by adjusting the time needed to distribute a Bell pair amongst two neighboring qubits to
\begin{equation}
 T_0=\frac{\beta L_0}{c}+T_M, \label{eq:T0}
\end{equation}
where $\beta$ is a factor depending on the position of the source which we choose to be $1$ and $c=2\times 10^{5}$ is the speed of light in the fiber. Apart from this change we use the formulas derived in \cite{abruzzo}.
The total amount of qubits is $2^{\tilde{N}+\tilde{k}+1}$. Hence the costs of the original repeater read
\begin{equation}
 C'=\frac{2^{\tilde{N}+\tilde{k}+1}}{R_{QKD}^O L} \label{eq:costsoriginal}.
\end{equation}
In the considered parameter regime the rate does not double when using distillation. It therefore never pays off to perform distillation with respect to the cost function $C'$, i.e. we set $\tilde{k}=0$. In our calculation we assume
\begin{equation}
 F_0=1-\frac{3}{4}f_G.
\end{equation}
This fidelity is obtained when using a gate to produce the initial Bell pair. Fig.~\ref{fig:originalrepeater} shows the cost comparison for a gate failure rate of $f_G=10^{-3}$ and three different measurement times $T_M=1\;\mathrm{\mu s},10\;\mathrm{\mu s},100\;\mathrm{\mu s}$.\\
One immediately sees that the costs of the one-way repeater scheme are proportional to the measurement time $T_M$. This is clear from the fact that this time is the only limiting factor in the repetition rate of this repeater. For the two-way repeater this is not the case. Decreasing the measurement time below approximately ten microseconds does not improve the costs, because then the communication time dominates the fundamental time (see Eq.~(\ref{eq:T0})) and becomes the limiting factor of the rate.\\
The sharp bends in the cost curve for the original repeater are due to the fact that \cite{Abruzzo13} considers only powers of two for the number of divisions of the transmission line. The straight line of the cost curve for the one-way repeater (over a large range of distances) shows that the costs per kilometer of this repeater using the Golay code increases polynomially with the total distance.\\
%The comparison of the repeater with the repeater scheme of~\cite{Muralidharan} is shown in Figure~??. For each distance $L$ the authors of \cite{Muralidharan} optimized over a class of error correction codes which are called quantum parity codes. The scheme uses two logical qubits per station.
\subsection{On the quality of some approximations}
\label{sec:approximations}
In the present paper we described the exact error analysis, mainly because the function $\Podd$ gives a convenient description of combined error rates. It is more readable than the evaluated polynomials, while the computational complexity is not an issue here. Nevertheless forward error correction requires a very low probability of operational errors of $\lesssim 10^{-2}$ in order for the processing of the qubits not to introduce more errors than are correctable. And thus it is reasonable to approximate the derived formulas for small error rates. On the other hand one usually considers the highest error rate that still allows to produce a secret key. This is the most interesting regime from a practical point of view due to the strong limitations of current technology. A similar effect arises from the use of the cost function as a figure of merit which punishes the use of resources and rewards e.g. higher losses in between the stations to some extent. Thus a critical verification of the accuracy of these approximations is advisable.\\
The first order estimates of $\Podd$(see Eqs.~(\ref{eq:PoddPN}) and (\ref{eq:Poddp})) are 
\begin{align}
 \Podd(P,N)=&N P + \mathcal{O}(P^2)\\
 \text{and}\hspace{1cm}\Podd(\vec{p})=& \sum_i (\vec{p})_i + \mathcal{O}((\vec{p})_i^2).
\end{align}
With these and $1-(1-f)^N\approx N f$ for small $f$ we find that (see Eqs.~(\ref{eq:fu}) and (\ref{eq:fn}))
\begin{align}
f_u\approx& \frac{3}{2} f_{P,u}+\frac{1}{2} f_{P,n} + \frac{3}{2} f_{G,u}+ f_{T,u}+\frac{1}{2} f_{M,u} \label{eq:fuapprox}\\
\text{and } f_n \approx & 2 f_{P,n}+3 f_{G,n}+2 f_{T,n}+2 f_{M,n}. \label{eq:fnapprox}
\end{align}
Because operational errors are small ($\lesssim 10^{-2}$), the second order contributions are even smaller and Eq.~(\ref{eq:fuapprox}) seems to be a good approximation. 
The losses however are typically bigger than ten percent (for repeater separations of $\gtrsim 1\,\mathrm{km}$, see Eq.~(\ref{eq:fTn})) so Eq.~(\ref{eq:fnapprox}) turns out to be a bad approximation, because second order contributions are not neglectable.\\
We use the Golay code to exemplify how the small inaccuracy of Eq.~(\ref{eq:fuapprox}) may become significant when the operational errors are near the maximally tolerable value in some situation and the number of repeater stations is large. Using the logical error rate given in Eq~(\ref{eq:golay}) one can calculate the cost $C'$. For a total distance of $L=600\;\mathrm{km}$, a gate error rate of $f_G=5\times 10^{-3}$ and $w=1500$ repeater stations it is $C' \approx 3464$ using Eq.~(\ref{eq:fu}) while it evaluates to $C'\approx 6500$ using the approximation of Eq.~(\ref{eq:fuapprox}). The discrepancy becomes even more obvious for slightly larger repeater separations. Setting $w=1400$ now $C'\approx 23448$ according to Eq.~(\ref{eq:fu}) while Eq.~(\ref{eq:fuapprox}) leads to a zero secret key rate (i.e. infinite costs).\\

\section{Generalization to the multipartite scenario}
\label{sec:multipartite}
\begin{figure}[tbp]%
 \begin{center}%
  \vspace{0pt}\hfill\subfigure[A mathematical graph.\label{fig:graphs:example}]{\includegraphics[scale=0.5]{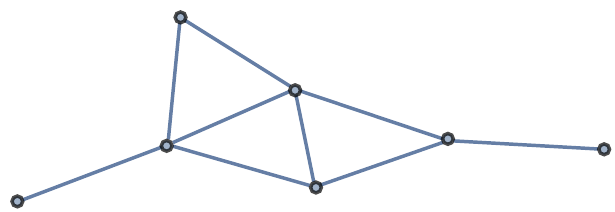}}\hfill%
  \subfigure[The line graph with six vertices.\label{fig:graphs:linegraph}]{\raisebox{5mm}{\includegraphics[scale=0.5]{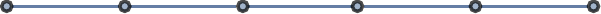}}}\hfill\vspace{0pt}%
 \end{center}%
 \caption{Examples of graphs.}\label{fig:graphs}%
\end{figure}

\begin{figure}[htbp]
 \begin{center}
  \subfigure[The final graph state.\label{fig:network:finalstate}]{\includegraphics[width=0.49\linewidth]{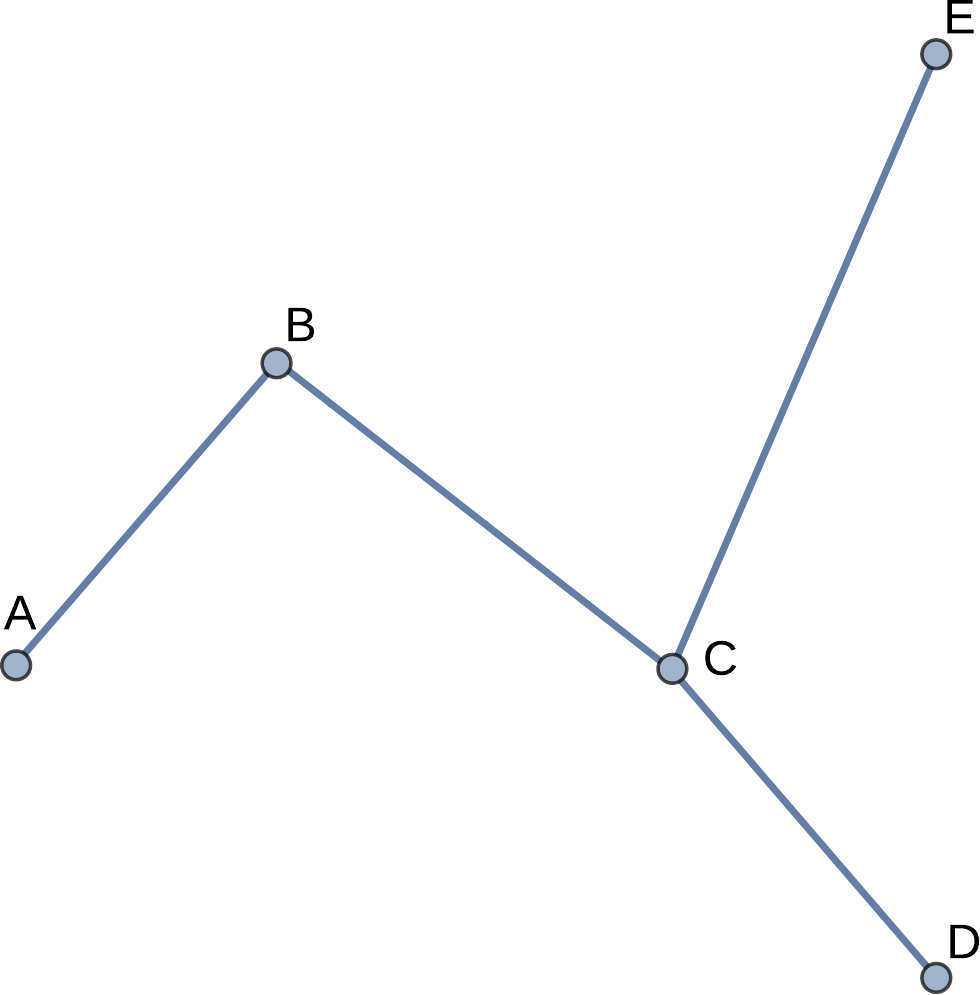}}
  %\subfigure[The network of intermediate repeater stations.\label{fig:network:stations}]{\includegraphics[scale=0.5]{network2}}\hfill
  \subfigure[The main stabilizer centered on C.\label{fig:network:stabilizers}]{\includegraphics[width=0.49\linewidth]{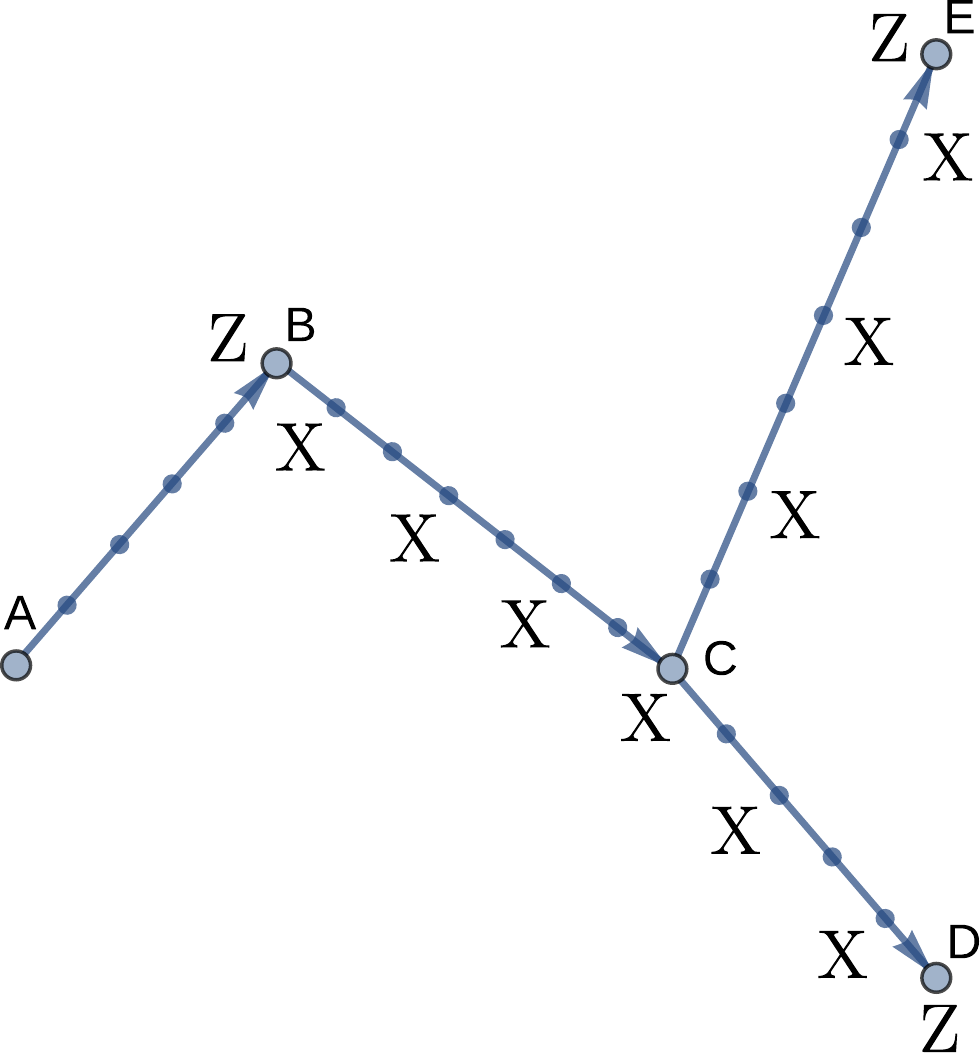}}
 \end{center}
 \caption{Example of a network of parties A to E.}\label{fig:network}
\end{figure}

We described in Section~\ref{sec:thecircuit} how the production of the final state can be understood in the stabilizer formalism. Measurements of the operators of the main stabilizers located on the intermediate qubits (i.e. all except the two of the parties) reduces the stabilizers to the stabilizers of the final state up to by-product operators. This procedure can be easily transferred to general graph states. We remind the reader that they are quantum states associated with mathematical graphs~\cite{Briegel01,Schlingemann01}. A Graph $G=(V,E)$ consists of a set of vertices $V$ and a set of edges $E\subset V\times V$, see Fig.~\ref{fig:graphs}~\subref{fig:graphs:example} for an example. We denote the number of vertices ($|V|$) by $N$.\\
The corresponding quantum state is the one stabilized by 
\begin{equation}
 g_i=X_i \prod_{\substack{j\\ (i,j)\in E}} Z_j,
\end{equation}
for all $i\in V$. One can arrive at these stabilizers by starting from $g_i=X_i$ (i.e. the state $\ket{+}^{\otimes N}$) and applying a $C_Z$ gate from qubit $i$ to qubit $j$ for all qubits $i<j$ with $(i,j)\in E$ (see Table~\ref{tab:errorpropagation}), i.e. for all edges in the graph. We thus note that the repeater circuit discussed in the previous sections (see Fig.~\ref{fig:basiccircuit}) creates a graph state where $E=\{(i,i+1)|1\leq i<N\}$. We call this graph a line graph (not to be confused with the line graph of a graph, i.e. the graph where vertices and edges exchange their role), see Fig.~\ref{fig:graphs}~\subref{fig:graphs:linegraph}.\\
Now the production/distribution of a general graph state is straight forward. To design the repeater network we start from the final graph and insert intermediate vertices for the repeater stations (see Fig.\ref{fig:network}). We insert an even number of repeater stations $w_{ij}$ on each edge $(i,j)\in E$, for simplicity. 
In analogy to the bipartite case the main stabilizer centered on some party is obtained by multiplication of the graph state generators centered on every second qubit until the neighboring parties are reached (with a $Z$-operator), see Fig.~\ref{fig:network}~\subref{fig:network:stabilizers}. On the added vertices the main stabilizer have the form of chains of $X$-operators (see also \cite{Wu15}). This ensures that the main stabilizers are transformed into the stabilizer generators $g_i$ of the final graph state by $X$-measurements on the repeater stations, i.e. the corresponding graph state is produced. The circuit is obtained again by noting that each edge of the graph corresponds to a $C_Z$ gate.\\
While the circuit of the repeater stations do not change compared to the bipartite case, the parties now apply more gates depending on the degree of their vertex (i.e. the number of edges at this position). Usually the number of repeater stations is much bigger than the number of parties for the error correction based scheme. One might therefore neglect the impact of the additional gates. Nevertheless they can be easily incorporated in Eqs.~(\ref{eq:fu}) and (\ref{eq:fn}) which become
\begin{equation}
 \begin{aligned}
f_{i,u}=&P_{\mathrm{odd}} 
\left(\left(P_{\mathrm{odd}}\left(\frac{f_{P,u}}{2},1+\deg^-(i)\right),\right.\right.\\
    &P_{\mathrm{odd}}\left(\frac{f_{P,n}+f_{P,u}}{2},\deg^+(i)\right),\\
    &P_{\mathrm{odd}}\left(\frac{f_{G,u}}{2},1+\deg(i)\right),\\
    &P_{\mathrm{odd}}\left(\left.\left.\frac{f_{T,u}}{2},1+\deg^-(i)\right),\frac{f_{M,u}}{2}\right)\right)
\end{aligned} \label{eq:fiu}\\
\end{equation}
and
\begin{equation}
\begin{aligned}
 f_{i,n} =&1-(1-f_{P,n})^{1+\deg^-(i)}(1-f_{G,n})^{1+\deg(i)}\\
 &(1-f_{T,n})^{1+\deg^-(i)}(1-f_{M,n})^{1+\deg^-(i)}, 
\end{aligned}\label{eq:fin}
\end{equation}
where $\deg(i)$, $\deg^-(i)$, and $\deg^+(i)$ are the degree, in-degree, and out-degree of vertex $i$, respectively. Here the direction of the edges corresponds to the direction of the transmission.\\
Note that local unitariy equivalence of graph states can be used to simplify the state distribution.
\section{Conclusions}
We described how quantum repeaters can be understood in the stabilizer formalism and how this formulation naturally leads to the description of general repeater networks.
Analyzing the error propagation in the circuit diagram leads to the error rates of the (physical) measurements on the repeater stations. To this end we identified all errors that may flip the measurement outcome at a specific repeater station in this circuit. It turns out that up to three repeater stations have to be considered in this calculation.\\
We calculated the secret key rate for a general CSS code given its logical error rate and exemplified this calculation with the Seven-Qubit-Steane code and the quantum Golay code. The comparison with the original quantum repeater scheme shows that the quantum Golay code is particularly resource efficient for large distances (and short measurement times of $\lesssim 10\;\mathrm{\mu s}$).\\
We investigated the quality of approximations of the physical error rates to the first order of the failure rates of the circuit elements (like gates) and found that these can be inaccurate in case of many repeater stations.\\
The repeater rate strongly depends on the abortion strategy, i.e. the set of error patterns on which one chooses to abort and restart the protocol. It is reasonable to abort on $d$ and more losses, where $d$ is the code distance.
\bibliographystyle{plain} 
%\bibliography{citations}

\appendix
\onecolumn
\section{The logical error rate of the Golay code}
\label{sec:golay}
We give the logical error rate of the decoder by M. Elia and G. Taricco~\cite{Elia95} for completeness. This decoder does not abort, so $P_{\mathrm{succ}}=1$. Note that we assume $\bar{f}_q\approx \frac{p_w}{2}$, where $p_w$ is the word error rate.\\
\begin{widetext}
\begin{equation}
\begin{aligned}
 \bar{f}_u(f_u,f_n)=& \frac{1}{2} \left(-\frac{f_n^{23}}{4096}+\frac{23 \left(f_n+f_u-1\right) f_n^{22}}{2048}-\frac{253
\left(f_n+f_u-1\right){}^2 f_n^{21}}{1024}+\frac{1771}{512} \left(f_n+f_u-1\right){}^3 f_n^{20}-\frac{8855}{256} \left(f_n+f_u-1\right){}^4
f_n^{19}\right.\\
&+\frac{33649}{128} \left(f_n+f_u-1\right){}^5 f_n^{18}-\frac{100947}{64} \left(f_n+f_u-1\right){}^6 f_n^{17}+\frac{245157}{32}
\left(f_n+f_u-1\right){}^7 f_n^{16}-30613 \left(f_n+f_u-1\right){}^8 f_n^{15}\\
&-\frac{253}{16} \left(f_n-1\right) \left(f_n+f_u-1\right){}^7 f_n^{15}+101200 \left(f_n+f_u-1\right){}^9 f_n^{14}\\
&+\frac{3795}{8} \left(f_n-1\right) \left(f_n+f_u-1\right){}^8 f_n^{14}-272734 \left(f_n+f_u-1\right){}^{10} f_n^{13}-\frac{26565}{4}
\left(f_n-1\right) \left(f_n+f_u-1\right){}^9 f_n^{13}\\
&+560924 \left(f_n+f_u-1\right){}^{11} f_n^{12}+\frac{115115}{2} \left(f_n-1\right) \left(f_n+f_u-1\right){}^{10} f_n^{12}-695520
\left(f_n+f_u-1\right){}^{12} f_n^{11}\\
&-319424 \left(f_n-1\right) \left(f_n+f_u-1\right){}^{11} f_n^{11}+\frac{8855}{2} \left(f_n+f_u-1\right){}^{11} \left(-f_n+2 f_u+1\right)
f_n^{11}\\
&+949256 \left(f_n-1\right) \left(f_n+f_u-1\right){}^{12} f_n^{10}-97405 \left(f_n+f_u-1\right){}^{12} \left(-f_n+2 f_u+1\right)
f_n^{10}\\
&+779240 \left(f_n+f_u-1\right){}^{13} \left(-f_n+2 f_u+1\right) f_n^9+18975 \left(f_n+f_u-1\right){}^{13} \left(-f_n+6 f_u+1\right)
f_n^9\\
&-485760 \left(f_n+f_u-1\right){}^{14} \left(-f_n+6 f_u+1\right) f_n^8-2277 \left(f_n+f_u-1\right){}^{14} \left(-f_n+14 f_u+1\right)
f_n^8\\
&+32384 \left(f_n+f_u-1\right){}^{15} \left(-f_n+14 f_u+1\right) f_n^7+\frac{253}{2} \left(f_n-1\right) \left(f_n+f_u-1\right){}^{14}
\left(-f_n+14 f_u+1\right) f_n^7\\
&+212520 \left(f_n+f_u-1\right){}^{14} \left(-\left(f_n-1\right){}^2+10 f_u \left(f_n-1\right)+8 f_u^2\right) f_n^7\\
&-100947 \left(f_n-1\right) \left(f_n+f_u-1\right){}^{15} \left(-f_n+14 f_u+1\right) f_n^6\\
&-28336 \left(f_n+f_u-1\right){}^{16} \left(-f_n+2 f_u+1\right) \left(-f_n+14 f_u+1\right) f_n^5\\
&-5313 \left(f_n+f_u-1\right){}^{16} \left(\left(f_n-1\right){}^2-15 f_u \left(f_n-1\right)+30 f_u^2\right) f_n^5\\
&+8855 \left(f_n+f_u-1\right){}^{17} \left(\left(f_n-1\right){}^2-17 f_u \left(f_n-1\right)+90 f_u^2\right) f_n^4\\
&-1771 \left(f_n+f_u-1\right){}^{17} \left(\left(f_n-1\right){}^3-17 f_u \left(f_n-1\right){}^2+138 f_u^2 \left(f_n-1\right)+96
f_u^3\right) f_n^3\\
&-253 \left(f_n+f_u-1\right){}^{18} \left(-\left(f_n-1\right){}^3+18 f_u \left(f_n-1\right){}^2-171 f_u^2 \left(f_n-1\right)+90
f_u^3\right) f_n^2\\
&+23 \left(f_n+f_u-1\right){}^{19} \left(-\left(f_n-1\right){}^3+19 f_u \left(f_n-1\right){}^2-190 f_u^2 \left(f_n-1\right)+560
f_u^3\right) f_n+\left(f_n+f_u-1\right){}^{23}\\
&\left.-23 f_u \left(f_n+f_u-1\right){}^{22}+253 f_u^2 \left(f_n+f_u-1\right){}^{21}-1771 f_u^3
\left(f_n+f_u-1\right){}^{20}+1\vphantom{-\frac{f_n^{23}}{4096}+\frac{23 \left(f_n+f_u-1\right) f_n^{22}}{2048}-\frac{253
\left(f_n+f_u-1\right){}^2 f_n^{21}}{1024}+\frac{1771}{512} \left(f_n+f_u-1\right){}^3 f_n^{20}-\frac{8855}{256} \left(f_n+f_u-1\right){}^4
f_n^{19}}\right)\\
\end{aligned}\label{eq:golay}
\end{equation}
\end{widetext}

\end{document}